\documentclass[prb,showpacs,showkeys,preprintnumbers,amsmath,amssymb,twocolumn]{revtex4-1}
\usepackage[T1]{fontenc} 
\usepackage{makeidx} 
\usepackage{graphicx} 
\usepackage{dcolumn} 
\usepackage{array} 
\usepackage{amssymb} 
\usepackage{amsmath}
\usepackage{textcomp}
\usepackage{rotating}
\usepackage{wasysym}
\usepackage{multirow}
\usepackage{subfigure}
\usepackage{eucal}
\usepackage{mathrsfs}
\usepackage{units}
\usepackage{rotating}
\usepackage[all]{xy}
\usepackage{float} 
\usepackage{amsmath} 
\usepackage{amsfonts} 
\usepackage{bm} 
\usepackage[amssymb]{SIunits}

\begin{document}

\title{Diamagnetic magnetocaloric effect due to a transversal oscillating magnetic field}

\author{M.S. Reis}\email{marior@if.uff.br}\affiliation{Instituto de F\'{i}sica, Universidade Federal Fluminense, Av. Gal. Milton Tavares de Souza s/n, 24210-346, Niter\'{o}i-RJ, Brasil}

\date{\today}

\begin{abstract}
The present Letter describes the magnetocaloric effect of a diamagnetic material with a magnetic field $B_\parallel$ along the $z$ axis and a transversal and oscillating field $B_\perp (\ll B_\parallel)$ parallel to the $x-y$ plane. We show that the magnetocaloric potentials due to a change in $B_\parallel$ is the same as those due to a change in the frequency of $B_\perp$. These results raise the possibility of building magnetocaloric devices without moving parts, since changing frequency is a simple electronic issue, while changing the field from permanent magnets depends on mechanical aspects.
\end{abstract}

\maketitle

The magnetocaloric effect (MCE) is an interesting property in which magnetic materials, under a magnetic field change, are able to exchange heat with a thermal reservoir (in an isothermal process), or even change its temperature (in an adiabatic process).\cite{reis2013fundamentals} This effect is completely analogous to the compression-expansion thermal-mechanical cycle, and therefore scientists hope to build a thermo-magnetic machine in the near future that can replace the current non-economical and non-environmentally friendly Freon-based refrigerators.\cite{tishin_book} Thus, for the last few decades scientists have explored possible applications of this effect to temperatures from the mK scale, e.g., an adiabatic demagnetization refrigerator,\cite{Timbie1990271} up to room temperature, e.g., air conditioners and industrial/domestic fridges.\cite{fridges,fridges2} Research on this issue thus has two main branches: device engineering\cite{yu2010review} and magnetic material optimization.\cite{gschneidnerjr2005recent,pecharsky2006advanced}

On the engineering side several devices have been proposed,\cite{yu2010review,tishin_book} but all of them need a magnetic field source such as a coil or a permanent magnet. Coils have serious disadvantages for achieving a magnetic field strong enough for the magnetocaloric effect to be useful (at least 2 Tesla\cite{tishin_book}), such as the need for liquid helium (for a superconducting coil), or huge water-cooled current sources (for a normal conducting coil). Thus, permanent magnets have been proposed and used in recent prototypes of magnetocaloric devices.\cite{zimm2006design} However, the magnetocaloric effect is due to a magnetic field change, so the use of a permanent magnet requires some part(s) of the device to move: either the refrigerant material or the permanent magnet.\cite{sarlah2006static}. It is therefore disadvantageous to use permanent magnets. As a consequence, most engineering efforts to design a new device focus on how to manage these moving parts.\cite{yu2010review}

Material scientists, on the other hand, focus on cooperative magnetic ordering, especially magneto-structural coupling because the magnetocaloric potentials are maximized near phase transitions.\cite{reis2013fundamentals} Thus, materials for magnetocaloric applications can only be ferro-, antiferro-, ferri- or paramagnetic,\cite{tishin_book} while diamagnetic materials have not been discussed until recently.\cite{reis2011oscillating,reis2012oscillating2,reis2012oscillating_apl,reis2013influence,reis2013electrocaloric,reisa2013oscillating}

From these last, thermodynamic properties oscillate because of the crossing of the Landau levels through the Fermi energy $\varepsilon_F$ of an electron gas (which models a diamagnetic material). A well-known example is the de Haas-van Alphen effect.\cite{greiner} Following this fact, we have recently described in detail the magnetocaloric properties of diamagnetic materials\cite{reis2011oscillating,reis2012oscillating2,reis2012oscillating_apl,reis2013influence,reis2013electrocaloric,reisa2013oscillating} and verified that they oscillate and can be either inverse or normal, depending on the value of the applied magnetic field change. Those results open the door for new applications as described in Ref. \onlinecite{reis2011oscillating}, namely in adiabatic demagnetization refrigerators and magnetic field sensors. This last proposal is based on the fact that this oscillating effect is sensitive to \emph{c.a.} 1 mT with a very high magnetic field in the background (\emph{c.a.}  10 T), at low temperatures, \emph{c.a.}  1 K.\cite{reis2011oscillating}

Following from the above, the aim of the present work is to show that it is possible to produce a magnetocaloric device without moving parts. For this purpose, we consider a diamagnetic material with a constant applied magnetic field $B_\parallel$ along the $z$ axis, and a transversal and oscillating magnetic field $B_\perp$ ($\ll B_\parallel$), with frequency $\omega$, parallel to the $x-y$ plane. We show therefore that the magnetocaloric effect due to a magnetic field change $\Delta B_\parallel:0\rightarrow B_\parallel$ is equivalent to the one with frequency change $\Delta\omega:\omega_\parallel\rightarrow 0$, where $\omega_\parallel$ is the resonance frequency of the oscillating magnetic field with the Landau levels of the diamagnetic material.

The starting point is the Hamiltonian of an electron under a magnetic field:
\begin{equation}\label{master}
\mathcal{H}=\frac{1}{2m}(\vec{p}+e\vec{A})^2,
\end{equation}
where $\vec{B}=\vec{\nabla}\times\vec{A}$. Considering first the magnetic field along the $z$ axis and, for simplicity, the Landau gauge, we have
\begin{equation}\label{bz_um}
\vec{B}=B_{\parallel}\;\hat{k}\;\;\;\Rightarrow\;\;\;\vec{A}=B_{\parallel}x\;\hat{j}.
\end{equation}
Substituting Eq. (\ref{bz_um}) into Eq. (\ref{master}) leads to a Schr\"{o}dinger equation in which the solution is a plane wave along the $z$ and $y$ axes, while the Landau states depending on $x$. The well-known eigenvalues are\cite{reis2013fundamentals}
\begin{equation}\label{aquii}
\epsilon=\left(j+\frac{1}{2}\right)\hbar\omega_{\parallel}+\frac{\hbar^2k_z^2}{2m}\;\;\;\textnormal{where}\;\;\;\omega_{\parallel}=\frac{eB_{\parallel}}{m},
\end{equation}
where $\hbar\omega_{\parallel}$ is the separation between Landau levels, $j$ is the Landau level index, and $k_z=2\pi n_z/L$ ($n_z=0,1,2,\cdots$) is related to the translational symmetry along the $z$ axis. Note that $\omega_\parallel$ is the cyclotron frequency for this electrons gas. 

Recently, the magnetocaloric potentials, i.e., magnetic entropy change \cite{reis2011oscillating} and adiabatic temperature change,\cite{reis2012oscillating2} of this diamagnetic system were evaluated. To obtain these, the entropy was first obtained via\cite{reis2011oscillating,reis2012oscillating2}
\begin{equation}\label{entrop}
S(T,B_{\parallel})=S(T,0)+S^o(T,B_{\parallel}),
\end{equation}
where the first term represents the zero-field contribution, while the second one oscillates depending on the reciprocal magnetic field $1/B_\parallel$. 

The magnetic entropy change is then\cite{reis2011oscillating}
\begin{align}
\Delta S(T,\Delta B)&=S(T,B_{\parallel})-S(T,0)\\\nonumber
&=S^o(T,B_{\parallel})\\\nonumber
&=-Nk_B\frac{3}{2}\left(\frac{1}{n+1/4}\right)^{3/2}\cos(n\pi)\;\mathcal{T}(x),
\end{align}
where
\begin{equation}
n=\frac{\varepsilon_F}{\mu_BB_{\parallel}}-\frac{1}{4},
\end{equation}
\begin{equation}
\mathcal{T}(x)=\frac{x\mathcal{L}(x)}{\sinh(x)},
\;\;\;\textnormal{and}\;\;\;\ 
\mathcal{L}(x)=\coth(x)-\frac{1}{x}
\end{equation}
is the Langevin function. In addition,
\begin{equation}
x=\pi^2t\left(n+\frac{1}{4}\right)
\end{equation}
and $t=k_BT/\varepsilon_F\ll 1$. Above, $\varepsilon_F$ stands for the Fermi energy of the system. This last condition is imposed on the evaluation of the grand partition function (see details in Ref. \onlinecite{reis2011oscillating}). The condition is valid because the Fermi energy of metals is thousands of kelvins, much higher than any laboratory temperature. Note that the magnetic entropy change oscillates because of the cosine term, and the oscillations depend on the reciprocal magnetic field $1/B_\parallel$. For a piece of gold ($\varepsilon_F=5.51$ eV), $n=10^4$ leads to $B_\parallel=9.516$ T and a normal MCE, while $n=10^4+1$ leads to $B_\parallel=9.515$ T and an inverse MCE. Note that only 1 mT is enough to change the MCE from normal to inverse. This effect is maximum at $x=1.6$ and therefore for $n=10^4$ and $T=1.04$ K.\cite{reis2011oscillating} At this temperature and magnetic field, the magnetic entropy change is c.a. $2\times 10^{-5}$ J/kg-K. The same was found for graphenes, and the entropy change is c.a. $5\times 10^{-2}$ J/kg-K at 10 T and 109 K.\cite{reis2011oscillating,reis2012oscillating2,reis2012oscillating_apl,reis2013influence,reis2013electrocaloric,reisa2013oscillating}

The adiabatic temperature change $\Delta T_{\Delta B}=T_f-T_i$, where the final state $f$ is the one with applied magnetic field $B_\parallel$ and the initial state $i$ has no applied field, was also evaluated.\cite{reis2012oscillating2} The condition to obtain $\Delta T_{\Delta B}$ is
\begin{equation}\label{cond}
S(T_i,0)=S(T_f,B_\parallel),
\end{equation}
where
\begin{equation}
S(T,0)=Nk_B\frac{\pi^2}{2}t
\end{equation}
and $S(T,B_\parallel)$ is given by Eq. (\ref{entrop}). After a few steps described in detail in Ref. 12, the adiabatic temperature change is obtained:
\begin{equation}
\Delta T_{\Delta B}=T_i\frac{\cos(n\pi)}{\sqrt{n}}.
\end{equation}

The present Letter thus goes further and adds an oscillating magnetic field transversal to the previous field, considering $B_\perp \ll B_\parallel$. In the laboratory reference frame, the total magnetic field applied to the electron gas is 
\begin{equation}\label{bz_bp}
\vec{B}_l^\omega=B_{\perp}\left[\cos(\omega t)\;\hat{i}+\sin(\omega t)\;\hat{j}\right]+B_{\parallel}\;\hat{k}.
\end{equation}
However, it is easier to treat this problem by changing the reference frame to one that rotates around the $z$ axis with angular frequency $-\omega$ and shares the origin with the laboratory frame. This procedure is fundamental to some resonant techniques such as nuclear magnetic resonance.\cite{guimaraaes1998magnetism} Thus, in the rotating frame, the magnetic field becomes
\begin{equation}\label{bz_bp}
\vec{B}_r^\omega=B_{\perp}\;\hat{i}+\left(B_{\parallel}-\omega/\gamma\right)\;\hat{k},
\end{equation}
where $\gamma=e/m$. Note the special condition for $\omega=\omega_{\parallel}=\gamma B_{\parallel}=eB_{\parallel}/m$ (see equation \ref{aquii}), i.e., the applied oscillating magnetic field is in resonance with the Landau levels. 

Using the Landau gauge as before, the vector potential becomes
\begin{equation}\label{bz_bp}
\vec{A}_r^\omega=\left(B_{\parallel}-\omega/\gamma\right)x\;\hat{j}+B_{\perp}y\;\hat{k}.
\end{equation}
Substituting Eq. (\ref{bz_bp}) into Eq. (\ref{master}) leads to a Schr\"{o}dinger equation with a plane wave along the $z$ axis as a solution, while the Landau levels depending on $x$ and $y$. Before delving deeper into this general case, it is useful to discuss the results of the Schr\"{o}dinger equation at the resonance, i.e., $\omega=\omega_{\parallel}$. For this particular case, the solutions are plane waves along the $z$ and $x$ axes, and the Landau states change to become dependent on $y$. The eigenvalues at the resonance are
\begin{equation}
\epsilon=\left(j+\frac{1}{2}\right)\hbar\omega_{\perp}+\frac{\hbar^2k_x^2}{2m}\;\;\;\textnormal{where}\;\;\;\omega_{\perp}=\frac{eB_{\perp}}{m},
\end{equation}
and $\hbar\omega_{\perp}$ is the new separation between Landau levels. It is interesting to note that the Landau states without the transversal oscillating magnetic field depend on $x$ and then change to become dependent on $x$ and $y$ when the transversal magnetic field is turned on. Then, at the resonance, those states change again and become dependent on $y$.

Back to the case far from the resonance, the solution of the Schr\"{o}dinger equation is more easily obtained if the magnetic field $\vec{B}_r^\omega$ is axial. For this we need to rotate the $z-x$ plane by an angle
\begin{equation}
\theta=\arctan\left(\frac{B_\perp}{B_{\parallel}-\omega/\gamma}\right).
\end{equation}
Thus, in this new frame, the applied magnetic field is
\begin{equation}\label{bz_bpp}
\vec{B}_a^\omega=\sqrt{B_{\perp}^2+\left(B_{\parallel}-\omega/\gamma\right)^2}\;\hat{k},
\end{equation}
and we obtain a problem with a well-known solution (described in detail in Refs. \onlinecite{reis2011oscillating} and \onlinecite{reis2012oscillating2}, as well as earlier in the present Letter). Thus, analogously to Eq. (\ref{entrop}), the magnetic entropy for the present case is
\begin{equation}
S(T,B_a^\omega)=S(T,0)+S^o(T,B_a^\omega).
\end{equation}

The standard MCE depends on a magnetic field change, and a magnetocaloric device therefore needs a moving part: either the permanent magnet or the magnetic material needs to move. The consequences of these moving parts are noise, energy consumption, complex design, and other undesired effects.

The main idea of the present Letter is therefore to produce a magnetic entropy change due to a change in the frequency of the transversal magnetic field, i.e., $\Delta\omega:\omega_\parallel\rightarrow 0$. Thus
\begin{align}
\Delta S(T,\Delta \omega)&=S(T,B_a^0)-S(T,B_a^{\omega_{\parallel}})\\\nonumber
&=S^o(T,B_a^0)-S^o(T,B_a^{\omega_{\parallel}}),
\end{align}
where $B_a^0=B_\parallel$, i.e., $B_a^0$ is the case without a transversal oscillating magnetic field, and $B_a^{\omega_\parallel}=B_\perp\ll B_\parallel$, i.e., $B_a^{\omega_\parallel}$ is the case with a transversal oscillating magnetic field at resonance $\omega=\omega_\parallel$. Remember, this resonance occurs when the oscillating magnetic field matches the Landau levels. However, $B_\perp$ is of the order of a few mT, while $B_\parallel$ is of the order of 10 T. Thus, $S^o(T,B_\parallel)\gg S^o(T,B_\perp)\rightarrow 0$, and 
\begin{equation}
\Delta S(T,\Delta\omega)=S^o(T,B_\parallel)=\Delta S(T,\Delta B).
\end{equation}
The above equation is our aim: the magnetic entropy change due to a magnetic field change of $\Delta B:0\rightarrow B_\parallel\approx 10$ T is the same as the magnetic entropy change due to frequency change $\Delta\omega:\omega_\parallel\rightarrow 0$ under a constant applied magnetic field $B_\parallel$. Note $\omega_\parallel$ is of the order of THz.

Similarly as before, the adiabatic temperature change $\Delta T_{\Delta\omega}=T_f-T_i$ due to a frequency change can also be evaluated. For this case, the final state $f$ is the one without a transversal magnetic field, i.e., $\omega=0$, while the initial state $i$ is the one with a transversal oscillating magnetic field at the resonance, i.e., $\omega=\omega_\parallel$. The condition to obtain $\Delta T_{\Delta \omega}$ is
\begin{equation}
S(T_i,B_a^{\omega_\parallel})=S(T_f,B_a^0),
\end{equation}
which turns out to be
\begin{equation}
S(T_i,0)=S(T_f,B_\parallel).
\end{equation}
Note the above equation is the same as Eq. (\ref{cond}), and therefore
\begin{equation}
\Delta T_{\Delta B}=\Delta T_{\Delta\omega}.
\end{equation}
Analogously to the magnetic entropy change, the adiabatic temperature change due to a frequency change is the same as that due to a magnetic field change.

It is important to emphasize that the magnetic entropy change presented in this Letter comes from the grand partition function (see details in Ref. \onlinecite{reis2011oscillating}). The results from the usual Maxwell relation would be the same as those obtained here, since for the Maxwell relation we would need to obtain either magnetization or specific heat, which also comes from the grand partition function.

Summarizing, we considered diamagnetic materials with a magnetic field $B_\parallel$ along the $z$ axis and a transversal oscillating magnetic field $B_\perp$ parallel to the $x-y$ plane. We verified that the magnetocaloric potentials due to a magnetic field change are the same as those due to a frequency change (from zero to the resonance value). From the practical point of view, i.e., developing devices, this idea provides many benefits because frequency change depends only on electronic issues instead of mechanical designs. These results open the door for future magnetocaloric devices without moving parts, and analogous work needs to be done with ordered materials.

We acknowledge FAPERJ, CAPES, CNPq, and PROPPI-UFF for financial support. We are in debt with prof. Nivaldo Lemos, for a helpful discussion.

\end{document}